# Ensemble Forecasting of Power Quality Parameters

Max Domagk
Peter Feistel
Jan Meyer
TUD Dresden University of Technology
Dresden, Germany
max.domagk@tu-dresden.de

Marco Lindner
TransnetBW
Stuttgart, Germany

Jako Kilter
Tallinn University of Technology
Tallin, Estonia

*Abstract*— The growing integration of power electronic-based technologies has increased the necessity of power quality (PQ) monitoring in transmission systems. Although large datasets are collected by operators, their use is typically limited to compliance assessment. Medium- to long-term forecasting can enhance the value of these datasets by enabling proactive asset management and trend detection, despite challenges related to data heterogeneity and seasonality. This paper systematically evaluates individual and ensemble forecasting approaches for PQ parameters in transmission systems. More than 700 weekly time series from measurement campaigns in Germany and Estonia are analysed to assess various models and aggregation strategies within a structured ensemble framework. The results show that ensemble forecasts consistently outperform individual models in terms of accuracy and robustness, achieving significant improvements over seasonal naive benchmarks and the best-performing single models. Ensemble forecasting is therefore confirmed as a robust and scalable approach for long-term PQ prediction in transmission systems.

## I. Introduction

The proliferation of inverter-based technologies, such as wind and solar generation, BESS, HVDC systems, and FACTS devices, has increased the importance of power quality (PQ) in transmission networks. Consequently, transmission system operators (TSOs) have intensified PQ monitoring in high- and extra-high-voltage grids. To date, analyses of these data have largely been confined to compliance assessments against planning levels. However, accurate medium- to long-term forecasting of PQ parameters over horizons of several months would enable early detection of adverse trends and potential threshold exceedances. Such forecasts support proactive maintenance planning and informed investment decisions,

Despite this relevance, research on medium- to long-term forecasting of PQ parameters at the transmission level remains limited. Effective forecasting of PQ time series requires several key properties. Robustness is essential to cope with diverse PQ parameters and heterogeneous data sources, while computational efficiency is crucial when handling large collections of time series. Moreover, beyond point accuracy, forecast reliability must be considered, as individual models inevitably exhibit structural limitations. In this context, ensembles of multiple forecasting models offer a promising means to improve robustness and mitigate model uncertainty [1].

This paper addresses these challenges through a statistical analysis of historical PQ data. It first introduces forecasting models suitable for medium- to long-term prediction of PQ parameters, followed by the development of a systematic ensemble forecasting framework. The performance of individual and ensemble models is then evaluated and compared to validate the robustness and practical usefulness of ensemble approaches for PQ forecasting.

## II. Data Characteristics and Preprocessing

### A. Power Quality Parameter Variations

PQ parameters vary over time due to the influence of many different factors, mainly related to the electrical environment. Generally, three different types of variations can be distinguished: short-, medium- and long-term variations [2]. Short-term variations are usually characterized by daily and weekly intervals which are mainly influenced by the consumer and generation behaviour. Certain consumer configurations show distinctly different daily patterns (e.g., working days vs. weekends).

Medium-term variations over periods of several months are related to more general changes in consumer and generation behaviour that are mostly related to seasonal effects. Earlier nightfall in winter months can lead to an increased use of equipment such as lighting, and consequently to higher emission of harmonics. The generated energy of PV systems is significantly higher during the summer months (longer days, higher angle of insolation).

This paper focuses on medium- to long-term variations, with long-term referring to time horizons of several months to one year, essential for reliable system performance forecasts.

### B. Dataset

The dataset originates from two measurement campaigns within the transmission system of a German and Estonian TSO. It contains 14 German and 13 Estonian measurements sites spanning different voltage levels: 110 kV (17 sites), 220 kV (2 sites), 330 kV (2 sites), and 380 kV (6 sites). The duration of the measurements is at least 3 years for all sites. Data acquisition employed fixed PQ instruments that comply with the IEC 61000-4-30 Class A standard.

The monitored PQ parameters include different voltage quality parameters: long-term flicker severity (Uplt), voltage unbalance (UNB), total harmonic distortion of voltage (Uthd) and individual harmonic voltages (U03, U05, U07, …, U15) recorded at a 10-minute aggregation interval.

### C. Data Preprocessing

The original 10-min measurements are first aggregated to weekly values by computing the $95^{th}$ percentile for each full calendar week (Monday to Sunday). A week is considered valid if at least 95 % of the 10-min measurements are available, allowing for up to 8.5 h of missing data per week in order to increase data availability.

Missing weeks or gaps within the resulting weekly time series occur frequently, for example due to maintenance outages. To maximize the number of usable time series, a up to 20 % of missing weeks is allowed. Gaps are filled using the most recent weekly 95th percentile available within the last 10 weeks.

The weekly values are then normalized by relating them to their respective planning levels, which provides the respective utilization. The applied planning levels follow national grid codes: in Germany, VDE-AR-N 4120 for high-voltage networks and VDE-AR-N 4130 for extra-high-voltage networks; in Estonia, planning levels specified by the transmission system operator, which are aligned with IEC TR 61000-3-6.

Following the complete preprocessing procedure, a total of 716 time series of 3 years are available for analysis. An example time series illustrating the utilization of the voltage unbalance is shown in Fig. 1.

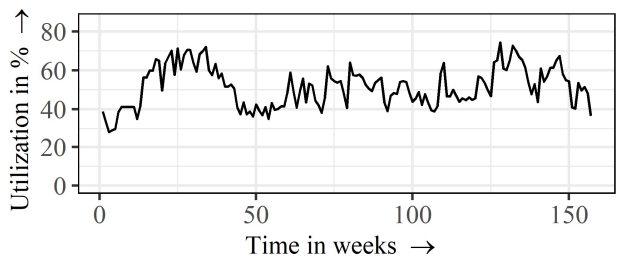

Fig. 1. Utilization of planning level for the voltage unbalance for a measurement site in the German transmission system within the 220 kV voltage level

## III. Forecasting Models

Forecast horizons range from very-short-term (seconds to minutes) to long-term (years to decades). Although machine and deep learning excel in short-term forecasting, their long-term efficacy is often hindered by a reliance on specific trend and periodicity structures [3]. Moreover, such "black box" approaches frequently lack explainability and demand significant computational resources [4]. Consequently, this study adopts interpretable models that decompose time series into trend, seasonal, and remainder components. Exponential Smoothing, ARIMA/SARIMA, and Prophet are utilized due to their computational efficiency, established validation, and robustness across diverse applications.

### A. Overview of Models

One category of forecasting approaches includes simple baseline methods. The Naive approach projects the last observed data point for all future values, while the Drift method extrapolates the average historical change. For seasonal time series, the Seasonal Naive (SNaive) approach equates future values to those from the corresponding period in the previous cycle and serves as the essential persistence baseline for benchmarking accuracy.

Another important class of forecasting techniques involves exponential smoothing methods. Exponential Smoothing (ES) generates forecasts as exponentially weighted averages of past observations [5], [6]. Extensions like Holt-Winters (HW) incorporate trend and seasonal components.

More sophisticated statistical approaches include ARIMA (Autoregressive Integrated Moving Average) models. The Box-Jenkins methodology models autocorrelation structures using non-seasonal (ARIMA) and seasonal (SARIMA) differencing [7]. These models are effective for data with strong autocorrelations and well-defined cycles.

A modern forecasting framework that has gained significant popularity is Prophet, which is an additive regression framework comprising trend, seasonality, and holiday components [8]. It utilizes piecewise-linear or logistic growth curves for trends and Fourier series for multiple seasonalities, offering flexibility for both periodic and non-periodic variations.

Finally, decomposition-based approaches offer an intuitive method for time series forecasting. STL decomposition (Seasonal and Trend decomposition using Loess) allows separate forecasting methods to be applied to isolated components [9]. For example, SNaive may be applied to the seasonal component while ARIMA is used for the seasonally adjusted series [10].

### B. Selected Models

In total, eight individual models are considered suitable for PQ parameter forecasting and are listed in Tab. I.

TABLE I. OVERVIEW OF INDIVIDUAL FORECASTING MODELS

| Model | Description |
| --- | --- |
| SNaive | Seasonal Naïve (benchmark) |
| HW | Holt-Winters Method |
| SARIMA | Seasonal ARIMA |
| Prophet | Additive model with trend and seasonality |
| STL-Drift | STL Decomposition with Drift Method |
| STL-ES | STL Decomposition with Exponential Smoothing |
| STL-Holt | STL Decomposition with Holt's Method |
| STL-ARIMA | STL Decomposition with ARIMA Method |

For the voltage unbalance utilization shown in Fig. 1, three representative model forecasts are exemplarily illustrated in Fig. 2. The three-year dataset is partitioned into a 105-week training set and a 52-week test set for evaluation. The HW model slightly overforecasts, while the Prophet and SNaive model slightly underforecast in this example.

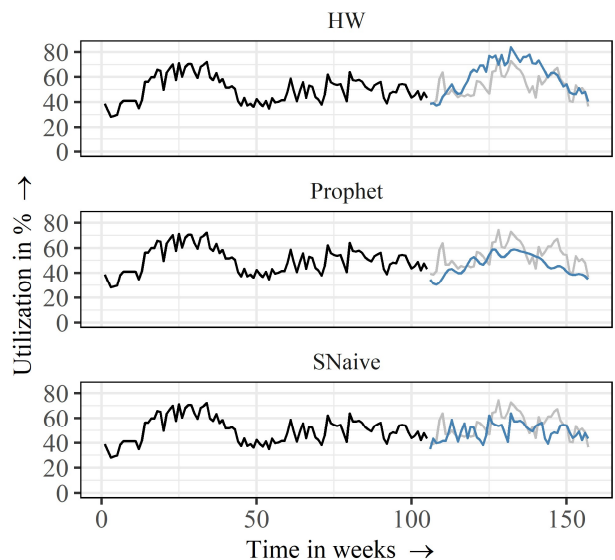

Fig. 2. Example forecast for the utilization of the voltage unbalance from Fig. 1 using three different models; training set (105 weeks) in *black*; test set in *grey* (52-week horizon) and model forecasts in *blue*

## IV. ENSEMBLE FORECASTING

### A. Ensemble Forecasting Framework

Ensemble forecasting is based on the generation of individual (base) forecasts followed by their combination. The process typically consists of two stages: a forecasting stage, in which multiple individual forecasts are produced from a given time series, and a combination stage, in which these forecasts are aggregated using a designated combination function to form a forecasting ensemble. The methods used to generate the individual forecasts, as well as the strategies for combining them, may vary substantially. Forecasts ensembles can be generated using approaches such as bootstrapping, hierarchical forecasting, or independent, non-interfering models [1].

This paper focuses exclusively on the third category, considering only ensembles derived from separate, non-interfering forecasting models.

### B. Ensemble Models

Using the eight individual forecasting models listed in Tab. I, ensemble models of varying sizes can be constructed, from combinations of two models up to all eight. All 247 possible configurations are considered, and each ensemble is assigned a short name in which the letter indicates the ensemble size (B = 2 models, C = 3 models, …, H = 8 models) and the number specifies its sequence within that group. For example, B01-B28 correspond to ensembles of 2 models, C01-C56 to ensembles of 3 models, and H01 represents the ensemble of all 8 models.

### C. Forecast Combination Methods

Forecast combination has a long history of improving forecasting accuracy and reducing model-selection risk, with extensive empirical evidence showing that ensembles often outperform their individual components [11]. The equal-weight mean is widely regarded as a robust baseline, effectively reducing variance and performing well across diverse time series. Median combinations provide an alternative that is less sensitive to outliers, although comparisons between mean and median results are mixed and largely data-dependent [1].

Weighted ensembles aim to give greater influence to better-performing models, for example by assigning weights based on recent model accuracy or relative performance [1]. In this work, four combination methods are considered and applied to all 247 ensemble configurations:

- **Mean** (equal-weight averaging of individual forecasts)
- **Median** (pointwise median of individual forecasts)
- **sMAPE** (accuracy-weighted ensemble based on mean model performance)
- **Rank** (rank-weighted ensemble based on mean model performance order)

The construction of the not equal weighted ensemble forecasts $\hat{y}_{\text{ens},T+h|T}$ using sMAPE and Rank follows a linear combination of $M$ individual model forecasts $\hat{y}_{i,T+h|T}$:

$$\hat{y}_{\text{ens},T+h|T} = \sum_{i=1}^{M} w_i \cdot \hat{y}_{i,T+h|T} \quad (1)$$

To ensure that models with higher accuracy (lower error or lower rank) receive greater weight, the weights $w_i$ are defined by the reciprocal of a performance metric $\phi_i$ normalized across all models:

$$w_i = \phi_i^{-1} / \sum_{j=1}^{M} \phi_j^{-1} \quad (2)$$

The input $\phi_i$ represents either the mean sMAPE or the mean rank of model $i$, as determined during the performance evaluation procedure.

An example forecast of ensemble C01 is shown in Fig. 3, generated by using the mean of the point forecasts of the three individual models from Fig. 2.

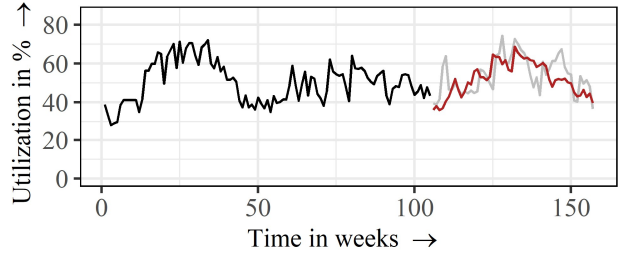

Fig. 3. Ensemble forecast for the utilization of the voltage unbalance using three different models from Fig. 2; training set (105 weeks) in *black*; test set in *grey* (52-week horizon) and ensemble forecast in *red*

## V. PERFORMANCE EVALUATION

### A. Performance Metrics

Point-forecast accuracy is evaluated using Mean Absolute Error (MAE) and symmetric Mean Absolute Percentage Error (sMAPE). For a forecast $\hat{y}_{T+h|T}$ and actual observed value $y_{T+h}$ over a horizon $H = 52$, sMAPE is defined as:

$$\text{sMAPE} = \frac{200}{H} \sum_{h=1}^{H} \frac{|y_{T+h} - \hat{y}_{T+h|T}|}{|y_{T+h}| + |\hat{y}_{T+h|T}|} \quad (3)$$

This metric is restricted to 0 % - 200 %, providing a balanced measure for evaluating forecast performance [12]. For this analysis, performance is classified as good below 10 %, acceptable up to 25 %, and poor or very poor thereafter.

### B. Performance Evaluation Procedure

Consistency is maintained by using a 105-week training period and a 52-week forecast horizon for all time series and models. Following the approach similar to [13], accuracy is assessed in three steps:

1. Calculation of accuracy (MAE and sMAPE) over the entire horizon for each model and time series.
2. Determination of the rank R(sMAPE) for each model per time series.
3. Calculation of the mean accuracy ($\overline{\text{MAE}}$ and $\overline{\text{sMAPE}}$) and mean rank ($\overline{\text{R(sMAPE)}}$) for each model across all 716 series.

These aggregated metrics ($\overline{\text{sMAPE}}$ and $\overline{\text{R(sMAPE)}}$) are subsequently used as the input for the construction of weighted ensembles as described in Section IV. C. To assess the performance relative to the baseline model the Benchmark Ratio (BR) is defined for any model $i$ as:

$$\text{BR}_i = \overline{\text{sMAPE}}_i / \overline{\text{sMAPE}}_{\text{SNaive}} \quad (4)$$

## C. Individual Model Performance

Tab. II provides an overview of the performance of the eight individual forecasting models across all 716 time series.

TABLE II. PERFORMANCE OF INDIVIDUAL FORECASTING MODELS

| # | Model | $\overline{\text{MAE}}$ | $\overline{\text{sMAPE}}$ | $\overline{R(\text{sMAPE})}$ | BR |
|---|---|---|---|---|---|
| 1 | STL-ARIMA | 3.9 % | 18.22 % | 3.34 | 0.873 |
| 2 | STL-ES | 3.93 % | 18.67 % | 3.28 | 0.894 |
| 3 | SARIMA | 4.08 % | 19.38 % | 4.08 | 0.928 |
| 4 | SNaive | 4.62 % | 20.88 % | 5.15 | 1 |
| 5 | Prophet | 4.51 % | 21.41 % | 4.3 | 1.025 |
| 6 | STL-Holt | 4.49 % | 21.53 % | 4.56 | 1.031 |
| 7 | STL-Drift | 5.34 % | 24.2 % | 5.39 | 1.159 |
| 8 | HW | 5.15 % | 25.03 % | 5.9 | 1.199 |

The results indicate that decomposition-based models showed clear dominance, with STL-ARIMA and STL-ETS being the top performers, achieving the lowest mean sMAPE (18.22 % and 18.67 %) and the best mean ranks. Only these STL variants and SARIMA successfully outperformed the SNaive benchmark (BR < 1), yielding accuracy gains of up to 12.7 %, while more complex models like Prophet and HW failed to surpass the baseline. These results validate the robustness of mean rank and the complementary of the metrics. For instance, while Prophet's sMAPE (21.41 %) was higher than SNaive's (20.88 %), their mean-rank ordering does not mirror this difference (4.56 vs. 5.15). This suggests that ordinal reliability can remain competitive even when error magnitudes vary, justifying the rank-based ensemble approach.

## D. Ensemble Model Performance

The analysis of 988 ensemble models (from 247 configurations using the four combination methods) shows significant improvement over individual models. As summarized in Tab. III, ensembles consistently achieve superior accuracy, with the top 100 models outperforming the best individual model (STL-ARIMA, sMAPE = 18.22 %). Notably, the best individual model does not rank within the top 100 ensemble models, highlighting the robust error-reduction capabilities of forecast combination. The D28 ensemble configuration (consisting of STL-ARIMA, STL-ES, SARIMA and SNaive) emerges as the overall winner, where the Median combination achieved the lowest sMAPE of 17.68 % and a Benchmark Ratio of 0.847. This represents an 2.6 % improvement over the best individual model and a 15.3 % improvement over the SNaive baseline.

TABLE III. PERFORMANCE OF TOP 100 ENSEMBLE MODELS

| # | Model | $\overline{\text{MAE}}$ | $\overline{\text{sMAPE}}$ | $\overline{R(\text{sMAPE})}$ | BR |
|---|---|---|---|---|---|
| 1 | D28 Median | 3.79 % | 17.68 % | 387.56 | 0.847 |
| 2 | D16 Median | 3.76 % | 17.74 % | 388.91 | 0.85 |
| 3 | D28 Rank | 3.81 % | 17.74 % | 372.36 | 0.85 |
| 4 | E23 Median | 3.81 % | 17.76 % | 359.49 | 0.851 |
| 5 | D28 sMAPE | 3.83 % | 17.77 % | 386.79 | 0.851 |
| … | … | … | … | … | … |
| 95 | D26 sMAPE | 3.9 % | 18.18 % | 410.57 | 0.871 |
| 96 | F18 sMAPE | 3.91 % | 18.18 % | 372.94 | 0.871 |
| 97 | E33 Median | 3.89 % | 18.19 % | 427.86 | 0.871 |
| 98 | E25 Mean | 3.9 % | 18.19 % | 393.94 | 0.871 |
| 99 | E01 Rank | 3.87 % | 18.19 % | 368.09 | 0.871 |
| 100 | D67 Rank | 3.86 % | 18.19 % | 370.8 | 0.871 |

As illustrated in Fig. 4, increasing the number of individual models in ensembles consistently reduces both the average sMAPE, improving accuracy, and the performance variance, enhancing robustness, with the top 100 ensembles achieving significantly higher accuracy compared to individual models. Furthermore, Fig. 5 reveals that the most successful ensembles frequently incorporate the five best individual models into groups of four or five. While the Median is the most frequent method in the top 100, the specific choice of aggregation is ultimately less critical than the composition and size of the model pool. Consequently, ensemble success appears to be predicated on the inclusion of diverse, well performing individual models.

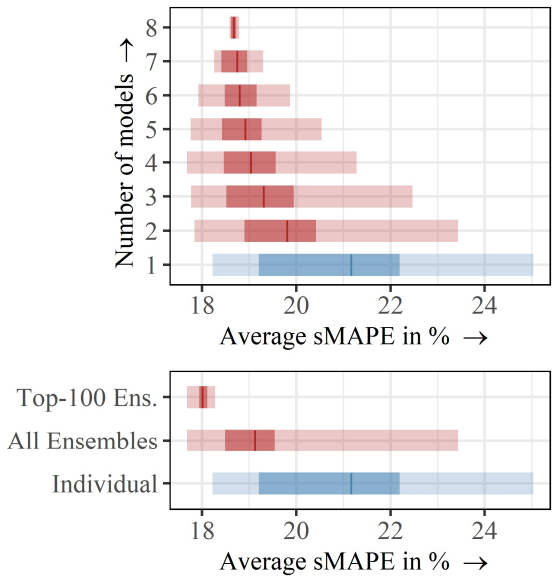

Fig. 4. Average sMAPE by ensemble size (*top*) and model set (*bottom*) with mean (*solid line*), 25th-75th percentile range (*dark bar*) and min-max range (*light bar*).

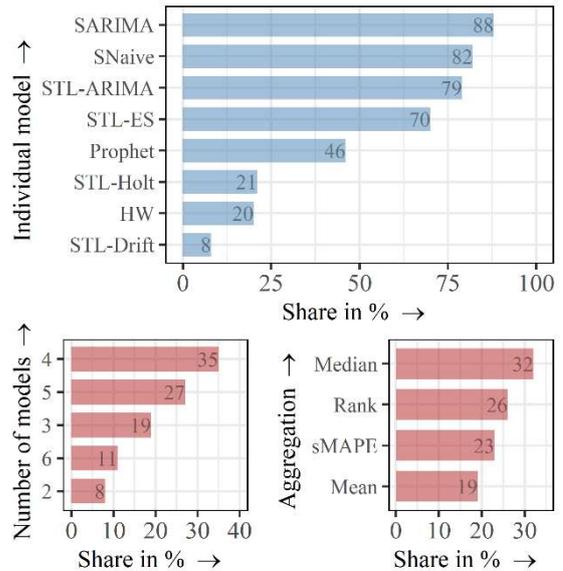

Fig. 5. Composition of top 100 ensemble models: individual model shares (*top*), ensemble size distribution (*bottom left*), and the utilization of various aggregation methods (*bottom right*)

## E. Comparison of the Best Individual and Ensemble Model

Fig. 6 provides illustrative examples of different forecasts for the best individual model and the best ensemble model.

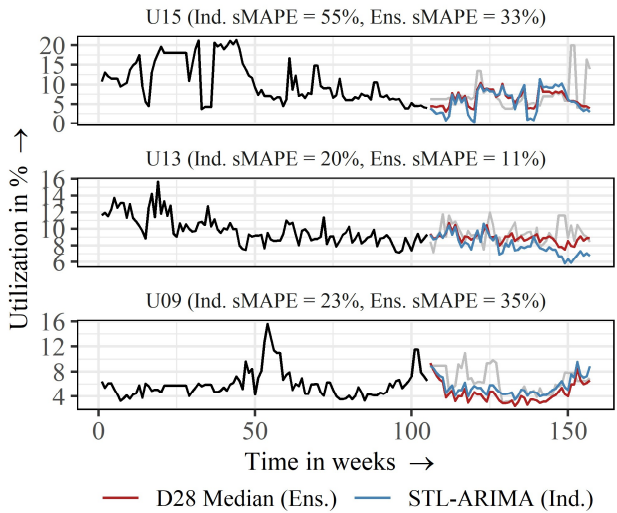

Fig. 6. Example forecasts of the best individual model (*blue*) and best ensemble (*red*): training set (105 weeks) in *black*; test set in *grey* (52-week horizon); From top to bottom: the ensemble performs substantially better, slightly better, and worse than the best individual model

To assess whether forecast combinations provide added value beyond careful model selection, their performances are compared in detail. The relative sMAPE improvement (cf. Fig. 7, left) shows that the best ensemble outperforms the best individual model in a majority of cases, achieving lower sMAPE in 62 % of all comparisons. When the ensemble is superior, the gains are systematic and practically relevant, with a median relative improvement of about 4 %. Although the individual model dominates in a minority of cases, the primary value of the ensemble lies in its robustness. By comparing individual versus ensemble sMAPE (Fig. 7, right), it can be seen that the ensemble stabilizes performance, preventing the extreme forecast failures often associated with selecting a single, sub-optimal model.

## VI. CONCLUSION

This study evaluated ensemble forecasting for medium- to long-term power quality prediction using transmission system data. While decomposition-based models such as STL-ARIMA and STL-ES exhibit strong standalone performance, no single model proves universally optimal across all time series. Ensemble methods effectively mitigate this model-selection risk by consistently improving forecast accuracy and reducing error variability. The results further indicate that ensemble performance depends primarily on the diversity of the individual models rather than on the complexity of the aggregation method, as simple mean or median combinations already yield robust and reliable improvements.

Consequently, ensemble forecasting offers a scalable and computationally efficient framework for transmission system operators to enhance trend identification and support infrastructure planning. Future research will focus on an application to distribution systems, extending it toward probabilistic forecasting, and developing asymmetric accuracy measures that prioritize forecast accuracy near power quality limits, where prediction errors have the greatest operational impact.

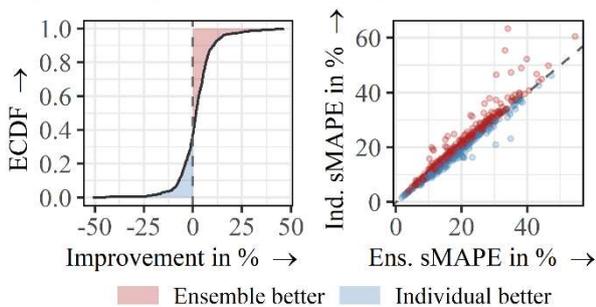

Fig. 7. Comparison of the best individual and the ensemble model: Empirical cumulative distribution function (ECDF) of relative sMAPE improvements (*left*) and scatter plot of individual versus ensemble sMAPE (*right*)

ACKNOWLEDGMENT

This work was supported by the German transmission system operator TransnetBW, the Estonian transmission system operator Elering AS, and the European Union, and was co-funded by the Estonian Ministry of Education and Science (Project No. TemTa-134).